\documentclass[aip,amsmath,amssymb,reprint,floatfix]{revtex4-1}

\usepackage{placeins}
\usepackage{graphicx}
\usepackage{dcolumn}
\usepackage{bm}

\usepackage[utf8]{inputenc}
\usepackage[T1]{fontenc}
\usepackage{mathptmx}
\usepackage{etoolbox}
\usepackage{xcolor}
\makeatletter
\def\@email#1#2{%
 \endgroup
 \patchcmd{\titleblock@produce}
  {\frontmatter@RRAPformat}
  {\frontmatter@RRAPformat{\produce@RRAP{*#1\href{mailto:#2}{#2}}}\frontmatter@RRAPformat}
  {}{}
}%
\makeatother
\begin{document}

\preprint{AIP/123-QED}

\title[]{\textbf{Shear-induced diffusivity in supercooled liquids}}

\author{Mangesh Bhendale}
\affiliation{Department of Chemical Engineering, Indian Institute of Technology, Kanpur, Uttar Pradesh 208016, India}

\author{Jayant K. Singh}
\affiliation{Department of Chemical Engineering, Indian Institute of Technology, Kanpur, Uttar Pradesh 208016, India}

\author{Alessio Zaccone}
\affiliation{Department of Physics ``A. Pontremoli'', University of Milan, via Celoria 16,
20133 Milan, Italy}
\email{alessio.zaccone@unimi.it}

\date{\today}

\begin{abstract}
    The Taylor-Aris theory of shear diffusion predicts that the effective diffusivity of a tracer molecule in a sheared liquid is enhanced by a term quadratic in the shear rate. In sheared supercooled liquids, instead, the observed enhancement is linear in the shear rate. This is a fundamental observation for the physics of nonequilibrium liquids. Here, we derive a formula for the effective molecular diffusivity in supercooled liquids under shear flow based on the underlying Smoluchowski equation with shear (Smoluchowski diffusion-convection equation) with an energy barrier due to the crowded energy landscape. The obtained formula recovers the effective diffusivity with a correction term linear in the shear rate, in reasonable agreement with results from numerical simulations of different liquids as well as with earlier experimental results on shear melting of colloidal glass. The theory predictions are compared with molecular simulations of supercooled water and supercooled Lennard-Jones liquids. The comparison suggests that the predicted enhancement of diffusivity is inversely proportional to temperature and directly proportional to the zero shear viscosity.     
\end{abstract}

\maketitle

%


\section{Introduction}
    The celebrated Taylor-Aris theory\cite{Taylor, Aris} of diffusion in a liquid undergoing shear flow (e.g. pipe flow) provides a foundation for understanding a variety of chemical and biochemical processes which occur in capillary flow as well as in industrial and environmental flows.
    The Taylor-Aris theory is based on solving the macroscopic diffusion-convection equation for a tracer particle in the absence of any conservative force-field or potential energy landscape (PEL). It provides a formula for the effective diffusivity enhanced by the shear flow given by:
\begin{equation}
    D_{\mathrm{eff}} = D \left( 1 + \frac{\mathit{Pe}^{2}}{48}\right)
\end{equation}
    where $Pe=R\bar{w}/D $ is the Peclet number, with $R$ the pipe radius, $\bar{w} \propto \dot{\gamma}$ is the average flow velocity in the pipe with $\dot{\gamma}$ the shear rate, and $D$ is the molecular diffusivity of the tracer particle in the absence of flow. This quadratic increase of the molecular diffusivity with the shear rate in normal liquids has been confirmed many times, both in experiments such as in NMR studies \cite{Codd} and in simulations \cite{Nanoscale,Louis}.

    In contrast with this result for a free diffusing molecule or particle in shear flow, the effective diffusivity measured experimentally or in numerical simulations in \emph{supercooled} liquids under shear flow, reads as \cite{Singh_PRL,Singh_pressure,Weitz_shear}
\begin{equation}
    D_{\mathrm{eff}} = D \left( 1 + c\dot{\gamma}\right) \label{new result}
\end{equation}
    for some constant $c$ independent of shear rate $\dot{\gamma}$. Hence, in supercooled liquids, the effective diffusivity is enhanced by a term that is linear in the shear rate, contrary to the Taylor-Aris result (valid for non-supercooled liquids), where the dependence on the shear rate is quadratic. This is a fundamentally unsolved problem in the statistical mechanics of nonequilibrium liquids\cite{Morriss_book} for which mostly numerical results are available.\cite{Kruger,Reichman} Seminal work by Schweizer and co-workers based on the microscopic Nonlinear Langevin Equation (NLE) led to unveiling a power-law dependence of the $\alpha$ relaxation time on the shear rate with an exponent $0.8$, for glassy hard sphere fluids. This prediction was confirmed within the more modern Elastically Collective Nonlinear Langevin Equation (ECNLE) theory in Ref.\cite{Schweizer_2020} and aligns with experiments on the flow of hard sphere colloids by Besseling et al.\cite{Besseling} Other theories and simulations \cite{Reichman} and experiments\cite{Weitz_shear} for glassy colloidal hard spheres found an exponent identically equal to 1.

    In the following, we provide the  physical derivation of Eq. \eqref{new result} based on the mathematical solution to the Smoluchowski diffusion-convection equation with a potential barrier representing the glassy cage in the supercooled liquid.

\section{Theory}
\begin{figure}[tbp]
    \centering
    \includegraphics[width=0.8\linewidth]{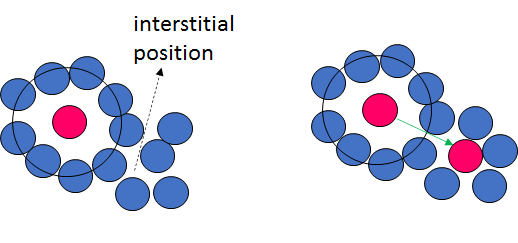}
    \caption{Schematic illustration of an event by which a particle abandons its original quasi-equilibrium position in the cage formed by its nearest-neighbours and jumps under the influence of thermal fluctuations to a new quasi-equilibrium position just outside the cage. The energy barrier $V_{\mathrm{max}}$ can be estimated as the elastic energy needed to accommodate the particle in the cavity, which for simplicity is taken to be spherical. This leads to a quantitative estimate of $V_{\mathrm{max}}$  according to e.g. \cite{Dyre}.}
    \label{cage}
\end{figure}

    The starting point is the Frenkel theory of diffusivity in a potential energy landscape. \cite{frenkel} In a crowded fluid, such as a supercooled liquid, the controlling process is the thermally activated hopping of a tagged molecule which escapes from the cage of its nearest-neighbors, as shown in Figure \ref{cage}.

From an energy-landscape perspective,\cite{debenedetti} this is a barrier-crossing process with a characteristic time-scale $\tau$, which, in the high-temperature liquid and for strong glasses, is an Arrhenius function of the local energy barrier $V_{\mathrm{max}}$, whereas it becomes a non-Arrhenius function for fragile liquids in the supercooled regime \cite{Angell,Dyre}. Within the mode-coupling theory (MCT) picture of supercooled liquids, the glassy cage is dynamical, and the particle hopping out of the cage is strictly tangled with the cage-relaxation. The above mentioned NLE and ECNLE theories go far beyond MCT to predict the dynamic free energy, energy barriers, activated hopping dynamics using Kramers theory, and effects of mechanical deformation.\cite{Saltzman_2008,Schweizer_2020} For practical purposes, a well-defined average barrier can be assumed, as customarily done in several contemporary theories of the glass transition.\cite{schweizer} We should, however, emphasize that the main result of our paper, i.e. the linear relationship between diffusivity coefficient and shear rate in supercooled liquids, is independent of the actual value of the energy barrier.

This energy barrier can be related, via the shoving model, to the elastic modulus \cite{Dyre,KSZ,puosi} and/or to the underlying glassy dynamics via microscopic NLE or ECNLE theory \cite{Saltzman_2008,Schweizer_2020,schweizer}.

According to Y. Frenkel, the diffusivity of a particle (atom, molecule) in an energy landscape is given by \cite{frenkel}:
\begin{equation}
    D=\frac{\delta^2}{6 \tau} \label{frenkel}
\end{equation}
where $\delta$ is the characteristic length-scale of the barrier-crossing process, typically of the order of the cage size (cfr. Figure \ref{cage}), hence 2-3 times the particle diameter. 

In the absence of shear flow, the hopping time scale for the particle to diffuse out of the cage is evaluated via the Kramers method, considering an energy barrier set by the cooperative slowing down predicted e.g. by mode-coupling theory.\cite{schweizer} In the presence of shear, the dynamics is described by the many-body Smoluchowski equation with shear \cite{fuchs}. Since we are interested in the shear-rate dependence, we assume the existence of a many-body potential barrier which arises from the glassy dynamics.\cite{schweizer}

Under these conditions, the dynamics of the tagged particle in the presence of drift terms, is governed by the Smoluchowski diffusion equation with shear~\cite{russel,Dhont,Zaccone2009,riva} for the probability density function (pdf) $\rho$ of finding the tagged particle at a position $\mathbf{r}$:
\begin{equation}
\frac{\partial \rho}{\partial t}+\nabla\cdot\left[-D_0 \nabla\rho+\mathbf{K}\rho\right]=0 \label{Smoluchowski}
\end{equation}
where $D_0$ is the single-particle diffusion coefficient in the high-temperature (not supercooled) liquid. In the above, $\mathbf{K}$ is the generalized drift, which contains the drift due to the PEL and that due to the shear flow \cite{Zaccone2009}. By the definition of the stationary current $\mathbf{J}$, we recover the continuity equation $\partial \rho/\partial t+\nabla \cdot \mathbf{J}=0$. At steady-state, the continuity equation dictates that the stationary current of probability density over a spherical surface is $J=4\pi r^{2}(-D_0 \partial\rho/\partial r+K_r \rho)$, where $K_r$ is the radial component of the drift field $\mathbf{K}$.

Let $\hat{r}$ be the unit vector measured from the center of the tagged particle along the outward trajectory. Clearly, only the current along this (positive) direction matters for the calculation of the barrier-crossing time $\tau$. The drift term in the presence of both an underlying PEL and an external flow reads as $K_r=-b(\partial V/\partial r)+b v_{r}$, where $b$ is the Stokes friction coefficient ($b = 6\pi \mu a$) with $a$ being particle radius, which is related to $D_0$ as, $D_0 = k_BT/b$. One should note that in the convection-diffusion equation studied by Taylor \cite{Taylor} and Aris \cite{Aris} the PEL term $-b(\partial V/\partial r)$ is absent.
Here $v_{r}\equiv \mathbf{v}\cdot\hat{r}$ is the radial component of the velocity due to the imposed shear flow.
    The spherically-averaged, radial current becomes \cite{Conchuir}
\begin{equation}
J=4\pi r^{2}\left(-D_0 \frac{\partial\rho}{\partial r}-b\frac{\partial V}{\partial r}\rho+b v_{r}\rho\right). \label{current}
\end{equation}
In polar coordinates, we integrate over all angles to find this radial current across a spherical cross section. However, only those regions of the solid angle where the flow drives the particle over the barrier of the glassy cage matter for the calculation of $\tau$. These regions correspond to regions of solid angle where $v_{r}$ is positive, whereas the regions where $v_{r}$ is negative do not contribute.\cite{Conchuir}
In particular, in the extensional regions of the solid angle (where the radial velocity component $v_{r} > 0$), particles can leave the cage because the relative velocity with respect to the center of the cage is positive and thus the particle is moved by the convective flow away from the center (outwards radially).
In the compressional regions (where $v_r < 0$), particles cannot leave the cage because the flow field, there, is pushing them towards the center of the cage (inwards radially).
Therefore, only the extensional regions (where $v_r > 0$) contribute to the net outflow; compressional regions (where $v_r < 0$) do not contribute.

Without loss of generality, we consider simple shear flow given by $\textbf{v}(x,y,z)=\dot{\gamma}[y,0,0]$ (other flow geometries can be implemented which is going to affect only a numerical prefactor in the final result). Under the assumption of weak-coupling between the flow field and the density field, $\rho(r)$ and $v_{r}$ are relatively uncorrelated over the solid angle (one should recall that $v_{r}$ also depends on the polar angle of the vector $\hat{r}$). This approximation has been checked by numerics in Ref. \cite{Zaccone2009} and shown to be able to yield reasonable results also for intense flows.
Hence, $\langle\rho v_{r}\rangle\approx \langle\rho\rangle \langle v_{r}\rangle$, where the $\langle...\rangle$ indicates the angular average restricted to the regions of the solid angle where the flow velocity acts as to move the particle at the center of the cage outwardly over the cage, i.e. where $v_r >0$. This is done always considering a spherical frame centered on the tagged particle at the center of the cage.
    In general, we have \cite{russel}
\begin{equation}
\begin{split}
\langle v_{r}\rangle&=\frac{1}{4\pi}\int_{\Omega}\dot{\gamma}r\sin^2 \theta \sin \phi \cos \phi\sin\theta d\theta d\phi\\
&=\frac{1}{3\pi}\dot{\gamma}r. \label{angular}
\end{split}
\end{equation}
To obtain the result in the second line, the angular integral is taken over the restricted set $\Omega$ of regions in the solid angle where the radial component of the flow velocity (and the associated drift) is positive along $\hat{r}$,\cite{Conchuir} thus pushing the particle away from the center of the cage over the barrier. These sectors of the solid angle where the relative velocity is directed outwardly are called "extensional sectors'', whereas the sectors of the solid angle where the relative velocity is directed inwardly are called "compressional sectors''. 
In particular, the sectors in the solid angle where $ v_{r} > 0$ are those in which 
\begin{equation}
    \sin \phi \cos \phi >0 
\end{equation}
where $\phi$ is the azimuthal angle, hence they correspond to regions of solid angle where
\begin{equation}
    \phi \in \{0, \pi/2\} \cup \{\pi, 3 \pi/2\} ~ ~ \forall \,\theta \in \{0,\pi\}.
\end{equation}
Obviously, this region of the solid angle is not restricted to the x-y plane but extends into the y-z plane as well.

For a different flow geometry, axisymmetric extensional flow, one would get $\langle v_{r}\rangle=\dot{\gamma}r/(3\sqrt{3})$ \cite{Conchuir}. It should be noted that, under a shear deformation $\gamma$, only particles in the extensional sectors of the solid angle can leave the cage. This effect cannot be cancelled out in the compressional sectors, because of excluded-volume. Hence, there is a net loss (outflow) of cage particles in the extensional sectors, which is not compensated by a corresponding influx in the compressional sectors, due to excluded volume. This issue has been discussed many times in the literature, e.g. with reference to Fig. 1 in Ref.\cite{Schall_2014}.

With an exact algebraic manipulation, we can rewrite Eq.\eqref{current} as~\cite{borkovec,Zaccone12}
\begin{equation}
J=-4\pi r^{2} D_0 e^{-V_{p}/k_{B}T}\frac{d}{dr}\left[e^{V_{p}/k_{B}T}\rho\right] \label{current new}
\end{equation}
where $V_{p}\equiv\int_{0}^{r} K_r ds$ is the primitive integral of the generalized drift $K_r$ introduced above.
Following the
Kramers' method \cite{Kramers}, we integrate Eq. \eqref{current new} between
$r^{*}$, a generic point near the cage center (corresponding to a point of minimum in the PEL $V(r)$), and $C$. Here $C$ is some point sufficiently away on the radial axis beyond the cage. Since the
probability density becomes much smaller at
$r=C$, we can express the steady current as
\begin{equation}
J=\frac{e^{V_{p}(r^{*})/k_{B}T}\rho(r^{*})}{a^{-2}\int_{r^{*}}^{C}\frac{e^{V_{\mathrm{eff}}(r)/k_{B}T}}{4
\pi D}dr} 
\label{eqmod3}
\end{equation}
where the effective potential is given by
\begin{equation}
V_{\mathrm{eff}}(r)\equiv V(r)-b\int_{0}^{r}\langle v_{r}\rangle s^2 ds-2k_{B}T\ln (r/a). \label{effective}
\end{equation}
This expression for the effective potential may have some similarities with the form of the nonequilibrium free energy formulated within NLE and ECNLE theories e.g. in Ref.\cite{Kobelev} although it was originally proposed in this form in the context of the theory of activated-rated processes in shear in Refs.\cite{Zaccone2009,Conchuir} This effective potential maps our 3D problem onto an effectively 1D problem but leaves the physics unaltered. The logarithmic term is necessary to recover the metric factor $r^{-2}$ in the integral of Eq. \eqref{eqmod3}, such that one can recover Eq. \eqref{current} upon going backwards in the transformation \cite{Ness}. The integral in Eq.\eqref{eqmod3} is indefinite, because it is the primitive integral (antiderivative), and the integration constant is chosen equal to zero such that we recover the case with no flow when $v_{r}=0$. 
The steady-state probability density inside the attractive well at the center of the cage is given
by the
stationary-state shear-distorted distribution $\rho_{st}$ by means of the quasi-steady state approximation in the well \cite{Conchuir},
$\rho_{st}(r)=\rho(r^{*})e^{-[V_{p}(r)-V_{p}(r^{*})]/k_{B}T}$ (this is simply the form which solves the steady-state time-independent limit of Eq. \eqref{Smoluchowski}).

Thus, the probability of finding the particle in the 3D well centered at the center of the cage is given by integrating the density over a spherical shell of this well,
\begin{equation}
\begin{split}
\rho_{st}=
\rho(r^{*})e^{V_{p}(r^{*})/k_{B}T}a^{2}\int_{A}^{B}e^{-V_{\mathrm{eff}}(r)/k_{B}T}
4\pi dr \, 
\end{split}
\label{eqmod4}
\end{equation}
where $A$ is a point slightly to the left of the PEL minimum (i.e. to the left of the cage center), and $B$ is a point slightly to the right.\cite{Kramers}
Upon taking $C\rightarrow \infty$, the mean first-passage time across the barrier is given by the Kramers theory \cite{Kramers,Nitzan} as $\tau=\rho_{st}/J$. Using the standard saddle-point method~\cite{Kramers} to approximate the integrals analytically to quadratic order both near the well bottom and near the barrier top in the integrals appearing in $\rho_{eq}$ and $J$, respectively, we obtain the time-scale for the shear-assisted crossing of the PEL cage barrier:
\begin{equation}
\tau=\frac{2\pi b \exp[\beta V_{\mathrm{eff}}(r_{\mathrm{max}})-\beta V_{\mathrm{eff}}(r_{\mathrm{min}})]}{\sqrt{-V''_{\mathrm{eff}}(r_{\mathrm{max}})V''_{\mathrm{eff}}(r_{\mathrm{min}})}}, \label{Kramers}
\end{equation}
where $r_{\mathrm{min}}$ and $r_{\mathrm{max}}$ represent the coordinates of the minimum and maximum in $V_{\mathrm{eff}}(r)$ and $\beta=1/(k_{B}T)$.

Upon substituting Eq. \eqref{angular} in Eq. \eqref{effective}, and then the latter in Eq. \eqref{Kramers}, we obtain:
\begin{equation}
\tau=\frac{2\pi b \exp[\beta V(r_{\mathrm{max}})-\beta V(r_{\mathrm{min}})]}{\sqrt{-V''(r_{\mathrm{max}})V''(r_{\mathrm{min}})}} e^{-\beta b \dot{\gamma} \Delta r^2/6\pi}, \label{Kramers2}
\end{equation}
where we separated the contribution due to the shear flow from that which survives in the limit of zero shear. Here, $\Delta r$ represents the spatial distance between the final position of the particle outside the cage and its original position at the center of the cage, hence $\Delta r \sim 2 \sigma = 4a$, with reference to the right panel in Figure \ref{cage}.
Since the shear velocity is linear in $r$, it does not change the location of the point of minimum and point of maximum, $r_{\mathrm{min}}$ and $r_{\mathrm{max}}$, respectively, of the PEL $V(r)$. Hence, $r_{\mathrm{min}}$ and $r_{\mathrm{max}}$ coincide with the minimum and maximum (separated by the cage barrier) of $V(r)$.
Furthermore, from the Stokes friction formula, we have: $b= 6 \pi \mu_0 a$, where $\mu_0$ is the liquid viscosity in the limit of zero shear rate and $a$ is the molecular radius. Upon replacing in the above formula, we finally obtain:
\begin{equation}
\tau=\tau_{\dot{\gamma}=0} e^{-\mu_0  \dot{\gamma} a \Delta r^2/k_B T}, \label{shear time}
\end{equation}
where we identified $\tau_{\dot{\gamma}=0} = \frac{2\pi b \exp[\beta V(r_{\mathrm{max}})-\beta V(r_{\mathrm{min}})]}{\sqrt{-V''(r_{\mathrm{max}})V''(r_{\mathrm{min}})}}$ as the barrier crossing time-scale in the absence of shear. We notice that the argument of the exponential in Eq. \eqref{shear time} is, correctly, dimensionless. 
Since $\Delta r \sim 4 a$, the argument of the exponential is a number very close to the particle Peclet number, i.e. $Pe \equiv 6 \pi \mu_0 \dot{\gamma} a^3 / k_B T$.
Upon substituting in Eq. \eqref{frenkel}, we get the following expression for the effective diffusivity:
\begin{equation}
    D_{\mathrm{eff}}=\frac{\delta^2}{6 \tau}=\frac{\delta^2}{6 \tau_{\dot{\gamma}=0}}e^{\mu_0  \dot{\gamma} a \Delta r^2/k_B T}=D e^{\mu_0  \dot{\gamma} a \Delta r^2/k_B T}.
\end{equation}
This is an important result, displaying an exponential dependence of the diffusivity on the shear rate. To our knowledge, this dependence has not been reported before although an exponential dependence on the shear rate has been predicted for the rate of thermally-activated processes in shear\cite{Zaccone2009} and even experimentally verified for the coagulation rate of charged colloids in shear flow.\cite{Zaccone_reaction,Dunstan} In what follows, we will show that this exponential can be very accurately approximated by a first-order Taylor expansion in $\dot{\gamma}$ because the prefactor $c$ multiplying $\dot{\gamma}$ in the exponential (Cfr. Eq.\eqref{new result}) is of the order of $10^{-10}$ seconds for molecular and atomic systems.\cite{Singh_PRL,Singh_pressure}

For molecular liquids, the molecule Peclet number is a small number, much smaller than 1, and therefore we can Taylor expand about $\dot{\gamma}=0$, to get 
\begin{equation}
    D_{\mathrm{eff}}=D\left(1 + \frac{\mu_0  a \Delta r^2}{k_B T}\dot{\gamma}\right) \label{final}
\end{equation}
which thus recovers the empirical form Eq. \eqref{new result} observed in simulations and experiments \cite{Singh_PRL,Singh_pressure,Weitz_shear} and thus identifies the prefactor as 
\begin{equation}
    c=\frac{\mu_0  a \Delta r^2}{k_B T}.
    \label{eq16}
\end{equation}
Equation \eqref{final} is the most important result of this paper, and provides the missing link between effective diffusivity, shear rate, viscosity, molecular size, and temperature in sheared supercooled liquids. It should be noted that this result is completely independent of the actual form of the average caging barrier, which may as well be highly dynamical and heterogeneous \cite{goetze}, and of the underlying PEL. This fact explains the observation of this law across many different systems, such as water, LJ and hard-sphere colloidal glasses.
\FloatBarrier

\begin{figure}[tbp]
    \includegraphics[width=0.9\linewidth]{mw-plots.eps}
    \caption{The variation in zero-shear diffusivity (a) and viscosity (b) of supercooled mW water with temperature, measured from NEMD simulations. The symbols represent the simulation data whereas the solid lines are linear fits to the simulated data.}
    \label{mw-plot}
\end{figure}

\begin{figure}[tbp]
    \includegraphics[width=0.9\linewidth]{lj-plots.eps}
    \caption{The variation in zero-shear diffusivity (a) and viscosity (b) of supercooled LJ particles with temperature, measured from NEMD simulations. The symbols represent the simulation data, whereas the solid lines are linear fits to the simulated data.}
    \label{lj-plot}
\end{figure}

\section{Molecular simulations}
    We have also performed numerical simulations to verify the above theoretical predictions. To this end, we performed nonequilibrium molecular dynamics (NEMD) simulations with two very different liquids, i.e., supercooled monoatomic water (mW)\cite{Molinero} and the Lennard-Jones (LJ) liquid. In the NEMD, the steady-state linear shear with Lees-Edwards boundary conditions (SLLOD) equations of motion were used.\cite{Daivis} 
    
    The mW model is capable of reproducing the thermodynamics, structure, phase transitions of water, and its interfacial properties.\cite{Molinero, Hudait-pccp-16, Moore-nature-11, moore-pccp-10, Mochizuki-jacs-17, Naullage-jacs-17, Naullage-jacs-18, Naullage-jacs-20, Naullage-jcp-20, Qiu-jacs-17, Qiu-jacs-19, Nguyen-jacs-15, Goswami-jcp-20, Singh_PRL, Metya-jacs-21, Singh_pressure} It mimics the hydrogen bonds using short-range two-body and three-body interactions and has extensively been used to study the interactions between ice and supercooled water in different supercooled conditions.\cite{Mochizuki-jacs-17, Naullage-jacs-17, Naullage-jacs-18, Naullage-jacs-20, Naullage-jcp-20, Qiu-jacs-17, Qiu-jacs-19, Nguyen-jacs-15, Metya-jppc-18, Metya-pccp-16, Goswami-jcp-20, Singh_PRL, Metya-jacs-21, Singh_pressure} For the mW model, simulations with $N=4096$ molecules were carried out in the temperature range $T=235 - 260$ K and in a broad range of shear rates $\dot{\gamma}=0.001 - 0.75$, in units of reciprocal simulation time. A multi-step equilibration procedure was applied in the supercooled regime following previous works.\cite{Singh_PRL,Singh_pressure} 
    To ensure the liquid state of the initial configuration, we first simulate 4096 mW water molecules at a temperature of 300 K for 25 ns, followed by cooling simulations with a cooling rate of 1 K/ns till a final temperature of 273 K is reached. The system is then further equilibrated for 5 ns at 273 K. The final configuration obtained after the 5 ns equilibration run is used as the initial configuration for running five independent NEMD simulations at the desired shear rate and supercooled temperature. The state of the mW beads as water and ice was distinguished using the CHILL+ algorithm. It identifies water molecules as liquid or ice (hexagonal and cubical ice polymorphs) based on the orientational order of a water molecule with its first four closest neighbours.
    
    We used the same protocol for the LJ liquid, with $N=4096$, temperature ($\epsilon/k_B$) in the range $0.55-0.7$ and shear rate in the range $0.001-0.5$ (both in LJ units, with $\epsilon$ the LJ energy scale and $\tau$ the LJ time scale). Similar to the case of mW water, the LJ liquid is initially equilibrated at T = 1 $\epsilon/k_B$ followed by a cooling down to T = 0.75 $\epsilon/k_B$. The final configuration from this equilibrated trajectory is used for NEMD shear simulations at the desired supercooling temperature.  

\begin{figure}[tbp]
    \includegraphics[width=\linewidth]{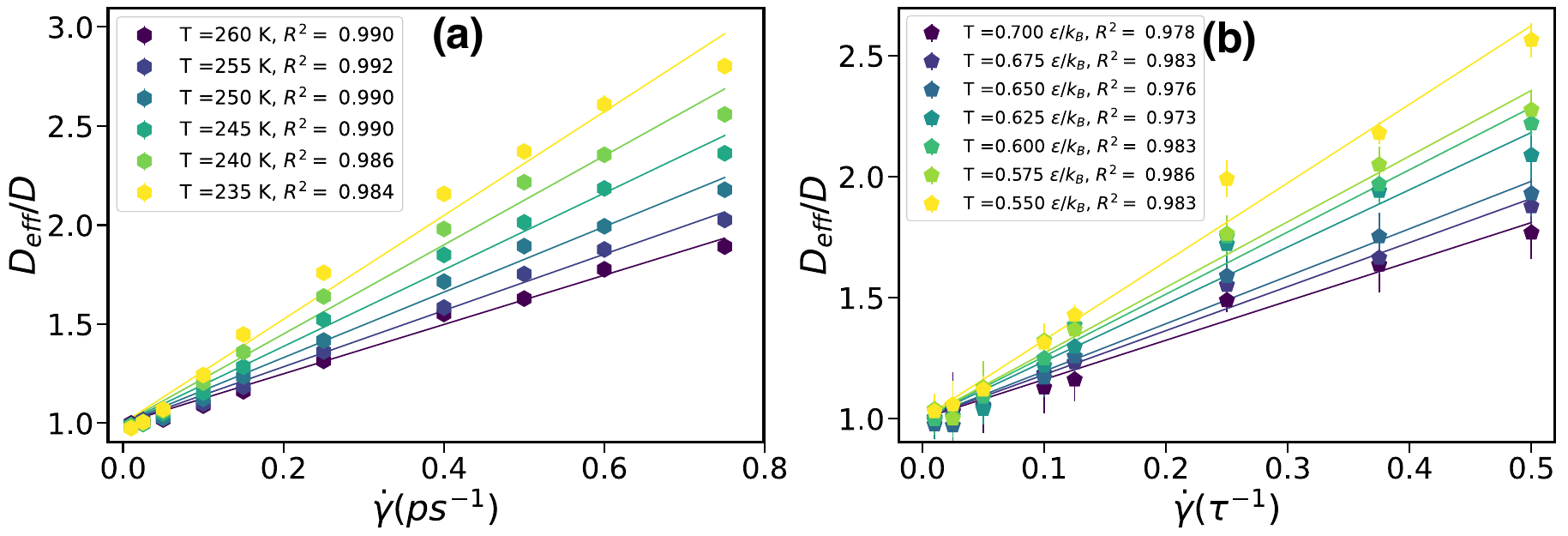}
    \caption{The plot of ($ D_{\mathrm{eff}}/D$) Vs shear, for supercooled mW water (a) and supercooled LJ particles (b), respectively, at different supercooled temperatures. The symbols represent the measured values from NEMD simulations, whereas the solid lines are best fits of the diffusivity ratio to the theoretically derived Eq. 15 in the paper.}
    \label{diffusivity-plot}
\end{figure}

\subsection{Shear diffusivity and viscosity} 
    The sheared transport properties, i.e., diffusivity and viscosity, were computed using the non-equilibrium molecular dynamics (NEMD)\cite{Todd-nemd} simulations with the LAMMPS (Large-scale Atomic/Molecular Massively Parallel Simulator) environment.\cite{lammps-Thompson-jcpc-22} The planar shear flow in the x-direction was simulated using the SLLOD\cite{Evans-pra-84, Ladd-molphy-84} equations of motion while keeping the temperature constant using a Nosé–Hoover thermostat with a relaxation time of 1 ps.\cite{nose-evans-jcp-1985} The 2D self-diffusion coefficient $D_{\mathrm{eff}}$ was calculated by taking ensemble averages over all molecules and time origins in the $y$ and $z$ dimensions for shear applied in the $xy$ plane. The viscosity was calculated by dividing the average stress by the shear rate $\dot{\gamma}$. Diffusivity and viscosity were calculated by averaging data from 5 independent trajectories.  Furthermore, the NEMD simulations were validated by analysing the velocity profile in the sheared $xy$-plane.

    The temperature dependence of the self-diffusion coefficient in the absence of shear (D) and of the zero shear viscosity ($\mu_{0}$) for supercooled (mW)\cite{Molinero} and LJ liquid are shown in Fig. \ref{mw-plot} and \ref{lj-plot}, respectively. In addition to being dependent on temperature, the transport properties also depend on shear.\cite{Singh_PRL} While shear flow enhances the molecular diffusivity of supercooled liquids, it also decreases their viscosity (shear thinning). The ratio of the effective molecular diffusivity enhanced by shear flow to its diffusivity in the absence of shear flow ($D_{\mathrm{eff}}/D$) is shown in Fig. \ref{diffusivity-plot}. Both supercooled mW water and supercooled LJ liquid exhibit an increase in the molecular diffusivity with an increase in the shear flow rate. The simulation results show that the enhancement in the molecular diffusivity due to shear flow diminishes as the degree of supercooling decreases. \par
    
\begin{figure}[tbp]
    \includegraphics[width=\linewidth]{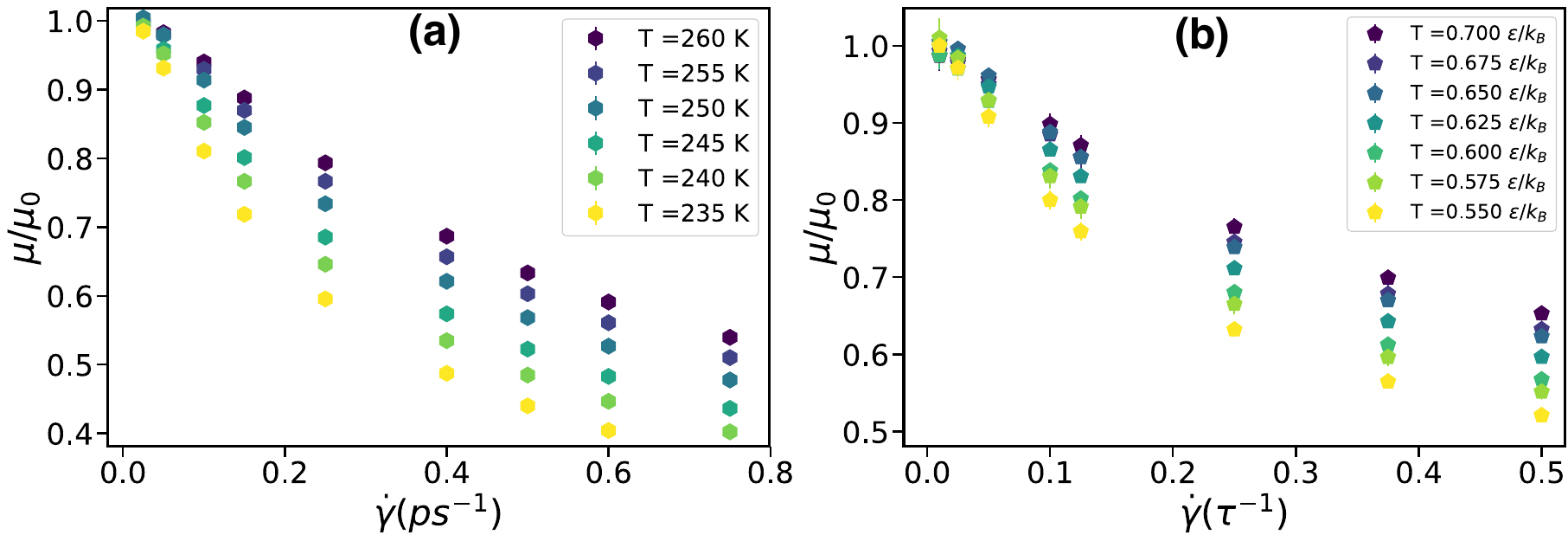}
    \caption{The ratio of viscosity ($ \mu_{\mathrm{eff}}/\mu_{0}$) in presence and absence of shear, for supercooled mW water (a) and LJ particles (b) at different shear rates, measured from NEMD simulations.}
    \label{viscosity-plot}
\end{figure}

    Similar to diffusivity, the viscosity of the supercooled liquid is also affected by flow. At very low shear rates, the viscosity remains constant, and the supercooled liquid exhibits Newtonian behavior. But at moderate and high shear rates, the viscosity decreases with an increase in shear rate, and this increase is non-monotonic in nature. Fig. \ref{viscosity-plot} shows the change in the ratio of viscosity ($\mu_{\mathrm{eff}}/\mu_{0}$) in presence and absence of shear, with a change in shear rate and degree of supercooling for supercooled mW water (Fig. \ref{viscosity-plot}(a)) and supercooled LJ liquid (Fig. \ref{viscosity-plot}(b)), respectively.\par

\begin{figure}[tbp]
    \includegraphics[width=\linewidth]{plot-delr.eps}
    \caption{The $\Delta r$ values from fitting $D_{\textrm{eff}}/D$ simulation data to Eq. 15, 
 for supercooled mW water (a) and LJ (b) particles, respectively.}
    \label{delr-plot}
\end{figure}

\subsection{Fitting of shear diffusivity}
    The shear diffusivity ratio ($D_{\text{eff}}/D$), obtained from the NEMD simulations was fitted to Eq. 15. The slope of the fit is the prefactor, $c$,  for supercooled mW water and LJ particles. The shear diffusivity ratio obtained from fitting is plotted against the shear diffusivity ratio obtained from NEMD simulations in Fig. \ref{diffusivity-plot}, as a function of shear rate. For fitting the simulation data, the values of molecular radius (a = $\sigma/2$) of mW water and LJ particles were considered to be $1.19625\times10^{-10}\ m$ and 0.5 $\sigma$, respectively.
    
    The values of 2D self-diffusion coefficient in the absence of shear (D), the zero-shear viscosity ($\mu_{0}$) used for fitting in Eq. 15, and the obtained prefactor $c$ for supercooled mW and LJ particles at different temperatures are reported in Table \ref{table1-mW} and \ref{table2-lj}. Later, the obtained $c$ values were used in Eq. 16, to calculate the $\Delta r$ for mW and LJ particles. The resulting $\Delta r$ values are plotted in Fig. \ref{delr-plot}. The $\Delta r$ values used for theoretical predictions in Fig. 2 and 3 are very close to the $\Delta r$ values resulting from NEMD data fitted to Eq. 16. While the $\Delta r$ values were assumed to be constant in the theoretical predictions, they are slightly lower when fitted to the NEMD simulation data. However, this decrease is within the error associated with the simulation data.\par

\begin{table}[th]
    \centering
    \begin{tabular}{cccc }
    \hline
        T & $D$ $(\times10^{-9} m^2/s)$ & $\mu_{0} (cP)$ & $c$ (ps)  \\
    \hline
        260	& 3.766	$\pm$ 0.048 & 0.478	$\pm$ 0.003  & 1.245 $\pm$ 0.028 \\
        255	& 3.487	$\pm$ 0.067 & 0.514	$\pm$ 0.001  & 1.421 $\pm$ 0.029 \\
        250	& 3.140	$\pm$ 0.053 & 0.558	$\pm$ 0.004  & 1.653 $\pm$ 0.038 \\
        245	& 2.830	$\pm$ 0.077 & 0.622	$\pm$ 0.005  & 1.935 $\pm$ 0.044 \\
        240	& 2.533	$\pm$ 0.095 & 0.687	$\pm$ 0.004  & 2.249 $\pm$ 0.059 \\
        235	& 2.262	$\pm$ 0.095 & 0.775	$\pm$ 0.007  & 2.620 $\pm$ 0.072 \\
    \hline
    \end{tabular}
    \caption{The values of the 2D self-diffusion coefficient in the absence of flow $D (\times10^{-9}\ m^{2}/s)$, of the zero-shear viscosity $\mu_{0}\ (cP)$, and of the prefactor c (ps) for supercooled mW water at different temperatures, T (K).}
    \label{table1-mW}
\end{table}

\begin{table}[th]
    \centering
    \begin{tabular}{ccccc }
    \hline
        T $(\varepsilon/k_B)$ & $D$ $(\sigma^2/\tau \times 10^{-2})$ & $\mu_{0} (\varepsilon\tau/\sigma^3)$ & $c$ $(\tau)$ \\
    \hline
        0.700 & 2.491 $\pm$ 0.104 & 4.500 $\pm$ 0.066  & 1.620 $\pm$ 0.065 \\
        0.675 & 2.268 $\pm$ 0.024 & 4.653 $\pm$ 0.091  & 1.819 $\pm$ 0.062 \\
        0.650 & 2.168 $\pm$ 0.071 & 4.730 $\pm$ 0.034  & 1.961 $\pm$ 0.083 \\
        0.625 & 1.964 $\pm$ 0.033 & 4.948 $\pm$ 0.005  & 2.364 $\pm$ 0.102 \\
        0.600 & 1.843 $\pm$ 0.033 & 5.210 $\pm$ 0.052  & 2.571 $\pm$ 0.087 \\
        0.575 & 1.700 $\pm$ 0.050 & 5.368 $\pm$ 0.101  & 2.712 $\pm$ 0.082 \\
        0.550 & 1.541 $\pm$ 0.028 & 5.687 $\pm$ 0.083  & 3.247 $\pm$ 0.109 \\
    \hline
    \end{tabular}
    \caption{The values of the 2D self-diffusion coefficient in absence of flow $D (\sigma^2/\tau)$, of the zero-shear viscosity $\mu_0 (\varepsilon\tau/\sigma^3)$, and of prefactor c $(\tau)$ for supercooled LJ particles at different temperatures, T $(\varepsilon/k_B)$.}
    \label{table2-lj}
\end{table}

\begin{figure}[tbp]
    \includegraphics[width=\linewidth]{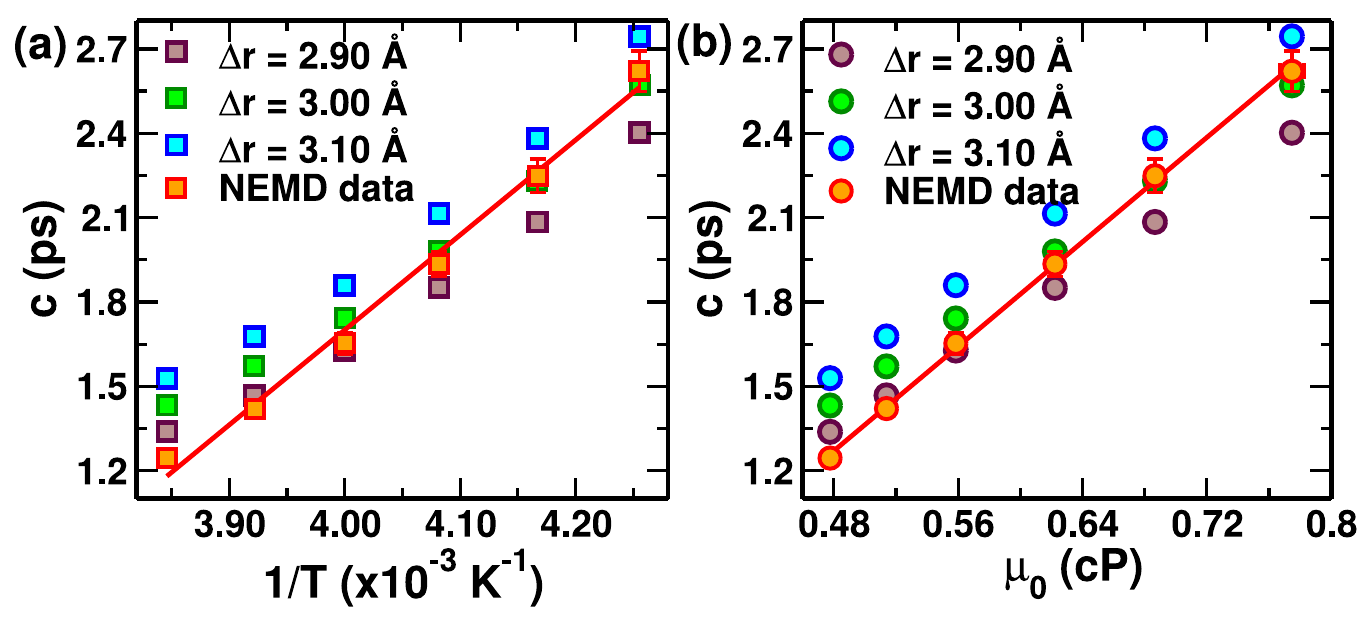}
    \caption{Symbols are the values of prefactor $c$ in  $D_{\mathrm{eff}} = D \left( 1 + c\dot{\gamma}\right)$ of supercooled mW water, plotted as a function of temperature $T$ in panel (a) and as a function of the zero shear viscosity $\mu_0$ in panel (b). The maroon, green, and blue symbols show the theoretical predictions using Eq. \eqref{eq16}, whereas the red symbols show the NEMD simulation data, respectively. Solid red lines are linear fit to the NEMD simulation data.}
    \label{fig2}
\end{figure}

\begin{figure}[tbp]
    \centering
    \includegraphics[width=\linewidth]{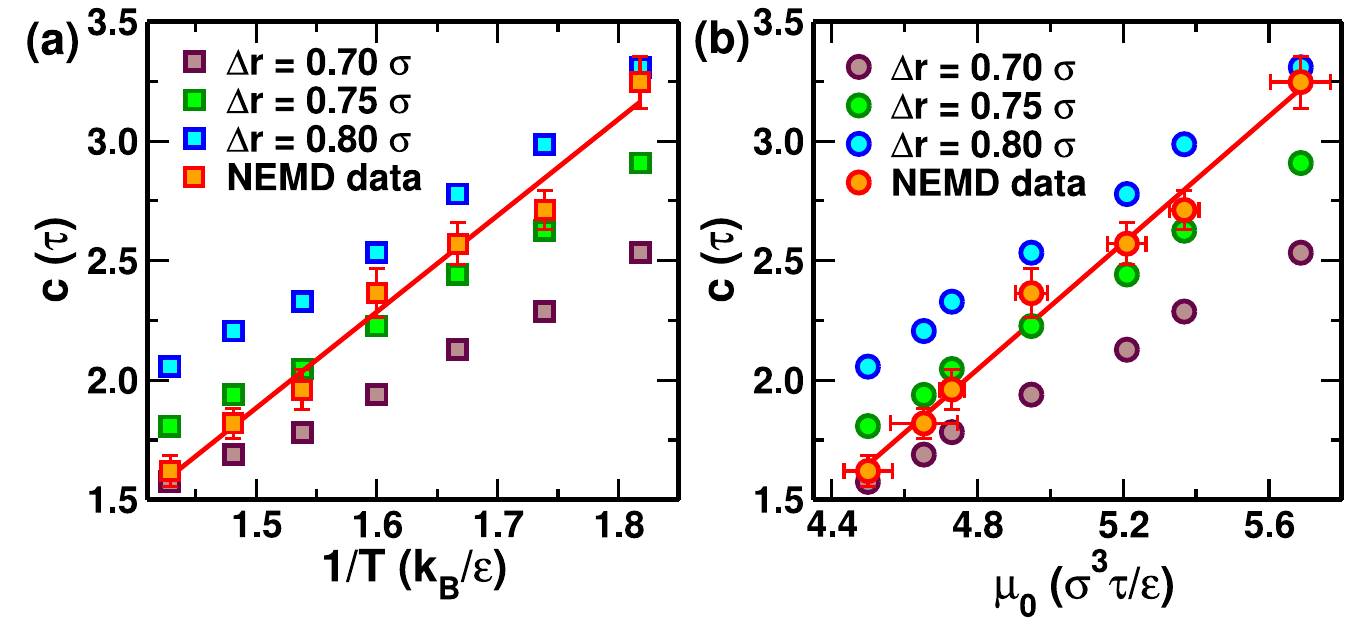}
    \caption{Symbols are the values of prefactor $c$ in  $D_{\mathrm{eff}} = D \left( 1 + c\dot{\gamma}\right)$ of supercooled LJ particles, plotted as a function of temperature $T$ in panel (a) and as a function of the zero shear viscosity $\mu_0$ in panel (b). The maroon, green, and blue symbols show the theoretical predictions using Eq. \eqref{eq16}, whereas the red symbols show the NEMD simulation data, respectively. Solid red lines are linear fit to the NEMD simulation data.}
    \label{fig3}
\end{figure}

\section{Comparison between theory and simulation data}
    We can now verify the above theoretical predictions in comparison with numerical simulations. For supercooled mW water, the effective diffusivity $D_{\mathrm{eff}}$ was found to perfectly follow the linear dependence on the shear rate $\dot{\gamma}$ given by Eq. \eqref{new result}, as shown in Figure \ref{diffusivity-plot}, as well as in previous works,\cite{Singh_PRL,Singh_pressure} and in the Supplementary Material therein. The NEMD simulation results however deviate slightly at low shear rates (Figure \ref{diffusivity-plot}) due to high errors associated with the predicted diffusivity ratios at low shear rates and the well-known time-scale bridging problem of slow dynamics in atomistic simulations. 

   From the linear fit, the coefficient $c$ was extracted for different conditions of temperature $T$ and of the zero shear viscosity $\mu_0$. The results are shown in Figure \ref{fig2} and supports the analytical result derived in Eq. \eqref{final}.

    Similar to mW water, the simulation data for supercooled LJ liquid was found to follow Eq. \eqref{new result} in previous works.\cite{Singh_PRL,Singh_pressure} Here again, we analysed the behaviour of the prefactor $c$ as a function of temperature and zero shear viscosity. The results are shown in Figure \ref{fig3}. Also in this case, the simulation results confirm the validity of Eq. \eqref{final} for both the predicted dependencies of $D_{\mathrm{eff}}$ on temperature $T$ and on the zero shear viscosity $\mu_0$.

\section{Conclusion}
    In summary, we have presented a mathematical theory of effective diffusion in shear flows valid for supercooled liquids, and we validated the theoretical predictions of the shear-induced self-diffusion enhancement by means of nonequilibrium molecular dynamics simulations for two very different fluids. 
    The enhancement of self-diffusion of a tracer molecule in an equilibrium (non-supercooled) liquid is well understood thanks to the Taylor-Aris dispersion theory, \cite{Taylor,Aris} which predicts an enhancement of diffusivity proportional to the square of the Peclet number. The Taylor-Aris theory is based on solving the governing convection-diffusion equation in the absence of any force fields to represent the local potential energy landscape. This assumption is no longer tenable in the supercooled regime, where molecular crowding leads to caging effects.\cite{Saltzman_2008,Schweizer_2020} These, in turn, represent an average energy barrier to the diffusive thermal hopping. \cite{debenedetti,schweizer} Hence, the problem has been reformulated in terms of the Smoluchowski diffusion equation with shear flow in an underlying (glassy) energy landscape. Importantly, the details of the barrier and of the glassy energy landscape, do not affect the final result. The equation has been solved analytically for the steady-state current using the Kramers' escape theory, and combining this result with Frenkel's theory of diffusivity leads to a shear-induced effective self-diffusion coefficient given by Eq. \eqref{final}. 

    Contrary to the Taylor-Aris result, the shear-induced enhancement of self-diffusion in the supercooled regime is now only linear in the shear rate, instead of quadratic. Furthermore, the enhancement is proportional to the zero shear viscosity and inversely proportional to temperature. Both these dependencies predicted by the theory are observed in nonequilbrium molecular simulations of supercooled water and of the supercooled Lennard-Jones liquid. This hints at the possible universality of the phenomenon and may explain previous experimental reports of shear diffusion in hard-sphere colloidal glass, where the linear scaling of the diffusivity with  $\dot{\gamma}$ was observed in Ref.\cite{Weitz_shear}. Deviations from the linearity relation are however possible:  for example, in a similar experiment on sheared colloids of a different lab\cite{Besseling}, a slightly smaller exponent was found, $D \sim \dot{\gamma}^{0.8}$. It should also be mentioned that a perfectly linear relation between structural relaxation time and shear rate was numerically predicted in 2D by mode-coupling theory in agreement with Brownian dynamics in Ref.\cite{Reichman}, while a shear thinning exponent $0.8$ between viscosity and shear rate was numerically observed in Ref.\cite{Furukawa}. Future extensions of this theory can address the cross-over from supercooled to equilibrium liquid upon increasing $T$, where the Kramers' escape theory has to be modified to recover free diffusion. \cite{Abkenar}

    Our theory provides insights into the mathematical form of the diffusivity in shear glassy systems, including the zero-order term. Indeed, our theory clarifies how the effective diffusivity reduces to the bare, unperturbed diffusivity when the shear rate goes to zero. Furthermore, our result provides new insights into the mathematical interrelation between diffusivity, zero-shear viscosity, and temperature, something that cannot be easily extracted from the numerical results scattered through various papers in the literature.
    
    All in all, given the technological importance of supercooled liquids, these results are expected to be beneficial for the quantitative modelling and rational control of mass transfer and  molecular and colloidal transport phenomena in a variety of physico-chemical systems.\cite{Bartosz2022,Wu1,Wu2,Robin}
                                        

\section{Supplementary Information}
    The Supporting Information is available free of charge at link.
    

\section{Authors' Contributions}
A.Z. conceptualised the project and developed the theory, whereas M.B. and J.K.S. carried out the computational work to validate the theoretical predictions. All authors contributed equally in writing the paper.

Mangesh Bhendale: Formal analysis (equal); Investigation (equal); Methodology (equal); Validation (equal); Visualization (lead); Writing – original draft (equal). 
Jayant K. Singh: Formal analysis (supporting); Funding acquisition (equal); Investigation (equal); Project administration (equal); Resources (equal); Supervision (equal); Validation (equal); Visualization (supporting); Writing – original draft (equal). Alessio Zaccone: Conceptualization (lead); Formal analysis (equal); Funding acquisition (equal); Investigation (equal); Methodology (equal); Project administration (equal); Resources (equal); Supervision (equal); Validation (supporting); Visualization (equal); Writing – original draft (equal).

\begin{acknowledgments}
A.Z. gratefully acknowledges funding from the European Union through Horizon Europe ERC Grant number: 101043968 ``Multimech'', and from US Army Research Office through contract nr. W911NF-22-2-0256. M.B. and J.K.S. thank the HPC and NSM supercomputing facilities of the Indian Institute of Technology, Kanpur, for providing the computational resources.
\end{acknowledgments}

\section*{Author Declarations}
The authors have no conflicts to disclose.

\section*{Data Availability}
The data that support the findings of this study are available from the corresponding author upon reasonable request.


\bibliography{bibliography}

\end{document}